# Frequency Diversity in Mode-Division Multiplexing Systems

Keang-Po Ho and Joseph M. Kahn

*Abstract*—In the regime of strong mode coupling, the modal gains and losses and the modal group delays of a multimode fiber are known to have well-defined statistical properties. In mode-division multiplexing, mode-dependent gains and losses are known to cause fluctuations in the channel capacity, so that the capacity at finite outage probability can be substantially lower than the average capacity. Mode-dependent gains and losses, when frequency-dependent, have a coherence bandwidth that is inversely proportional to the modal group delay spread. When mode-division-multiplexed signals occupy a bandwidth far larger than the coherence bandwidth, the mode-dependent gains and losses are averaged over frequency, causing the outage capacity to approach the average capacity. The difference between the average and outage capacities is found to be inversely proportional to the square-root of a diversity order that is given approximately by the ratio of the signal bandwidth to the coherence bandwidth.

## I. INTRODUCTION

ALTHOUGH multimode fiber (MMF) is used traditionally for short-reach links [1]-[3], the throughput of long-haul fiber systems can be increased, in principle, by mode-division multiplexing (MDM) in MMF [4]-[10]. Ideally, the channel capacity is directly proportional to the number of modes.

The modes in an MMF have slightly different group delays (GDs) [11] and potentially different losses. Manufacturing variations, bends, mechanical stresses, thermal gradients and other effects cause coupling between different modes [12][13]. The statistics of mode-dependent GDs and mode-dependent gains and losses (collectively referred to here as MDL) in the regime of strong mode coupling were studied by us recently [14][15]. MDL poses a fundamental limit to system performance [15][16]. The extreme case of high MDL is equivalent to a reduction in the number of modes, leading to a proportional reduction in channel capacity.

In wireless communications, multipath propagation causes frequency-selective fading of wideband signals [17]. Various forms of frequency diversity can be used to combat this effect. For example, using coded orthogonal frequency-division multiplexing (OFDM) [18][19], an error-correction code effectively averages over strong and weak subchannels. Alternatively, space-time codes can provide frequency diversity for OFDM signals [20][21], or for single-carrier signals [22][23].

In wireless communications, multipath channel models typically depend on many parameters including, but not limited to, the number of paths, the fading distribution for each path (e.g., Rician or Rayleigh), the delay spread, and the speed of the user. A single statistical model is often unable to include all important cases.

By contrast, for MMF in the strong-coupling regime, the channel statistics depend on only a few parameters, and simple statistical models are able to include all meaningful cases [14][15]. The statistics of the GDs depend only on the number of modes and the overall GD spread [14], while the statistics of the MDL, and thus the channel capacity, depend only on the number of modes and the overall MDL [15]. At any single frequency, the channel capacity is a random variable that depends on the specific realization of MDL, and the outage capacity may be significantly smaller than the average capacity [15][16].

The frequency dependence of MDL has a coherence bandwidth that should be inversely proportional to the GD spread. Likewise, the channel capacity has a coherence bandwidth that is also inversely proportional to the GD spread. If MDM signals occupy a bandwidth far larger than the coherence bandwidth of the capacity, because of statistical averaging, the outage capacity should approach the average capacity. These frequency diversity effects are studied numerically in this paper. For typical values of MDL and the signal-to-noise ratio (SNR), the coherence bandwidth of the capacity is found to be approximately equal to the reciprocal of the standard deviation (STD) of the GD, $\sigma_{gd}$. The difference between the average capacity and the outage capacity is found to decrease with the square-root of a diversity order that is given approximately by the ratio of the signal bandwidth to the coherence bandwidth of the capacity.

The remainder of this paper is organized as follows. Section II reviews the random matrix model from which the frequency-dependent GD and MDL statistics are derived, and presents the correlation coefficient of MDL as a function of frequency separation. Section III presents the correlation

Manuscript received August ??, 2011, revised ??, 2011. The research of JMK was supported in part by National Science Foundation Grant Number ECCS-1101905 and Corning, Inc.

K.-P. Ho is with Silicon Image, Sunnyvale, CA 94085 (Tel: +1-408-419-2023, Fax: +1-408-616-6399, e-mail: kpho@ieee.org).

J. M. Kahn is with E. L. Ginzton Laboratory, Department of Electrical Engineering, Stanford University, Stanford, CA 94305 (e-mail: jmk@ee.stanford.edu).

coefficient of channel capacity as a function of frequency separation, and describes how frequency diversity mitigates the frequency dependence of capacity. Sections IV and V provide discussion and conclusions, respectively. The Appendix describes method to compute diversity order directly from the frequency correlation coefficients.

## II. FREQUENCY-DEPENDENT PROPAGATION IN MMF

Long-haul MDM systems are expected to be in the strong-coupling regime, in which the overall fiber length is far longer than a correlation length over which the local eigenmodes can be considered constant [14][15]. In this regime, a fiber can be modeled as a concatenation of many independent sections.

### A. Random Matrix Model

An MMF is assumed to be composed of $K$ independent sections, each having length at least equal to the correlation length. Each section is modeled as a random matrix, as in [14]-[16]. This is an extension of the models used for polarization-mode dispersion or polarization-dependent loss in single-mode fiber [24][25]. The overall transfer matrix of an MMF comprising $K$ sections, as a function of angular frequency $\omega$, is:

$$\mathbf{M}^{(t)}(\omega) = \mathbf{M}^{(K)}(\omega) \cdots \mathbf{M}^{(2)}(\omega) \mathbf{M}^{(1)}(\omega) . \quad (1)$$

For an MMF supporting $D$ propagating modes[1] the matrix for the $k$th section is $\mathbf{M}^{(k)}(\omega)$, a $D \times D$ matrix that is the product of three $D \times D$ matrices:

$$\mathbf{M}^{(k)}(\omega) = \mathbf{V}^{(k)} \mathbf{\Lambda}^{(k)}(\omega) \mathbf{U}^{(k)*}, \quad k = 1, \ldots, K . \quad (2)$$

Here, * denotes Hermitian transpose, $\mathbf{U}^{(k)}$ and $\mathbf{V}^{(k)}$ are frequency-independent random unitary matrices representing modal coupling at the input and output of the section, respectively, and $\mathbf{\Lambda}^{(k)}(\omega)$ is a diagonal matrix representing modal propagation of the uncoupled modes in the $k$th section.

Including both MDL and modal dispersion, $\mathbf{\Lambda}^{(k)}(\omega)$ can be expressed as:

$$\mathbf{\Lambda}^{(k)}(\omega) = \mathrm{diag}\left[ e^{\frac{1}{2}g_1^{(k)} - j\omega\tau_1^{(k)}}, e^{\frac{1}{2}g_2^{(k)} - j\omega\tau_2^{(k)}}, \ldots, e^{\frac{1}{2}g_D^{(k)} - j\omega\tau_D^{(k)}} \right], \quad (3)$$

where, in the $k$th section, the vector $\mathbf{g}^{(k)} = \left(g_1^{(k)}, g_2^{(k)}, \ldots, g_D^{(k)}\right)$ describes the uncoupled MDL, and $\boldsymbol{\tau}^{(k)} = \left(\tau_1^{(k)}, \tau_2^{(k)}, \ldots, \tau_D^{(k)}\right)$ describes the uncoupled modal GDs.

Similar to multi-input multi-output (MIMO) wireless systems [26][27], at any single frequency, using singular value decomposition, the overall matrix $\mathbf{M}^{(t)}(\omega)$ can be decomposed into $D$ spatial channels:

$$\mathbf{M}^{(t)}(\omega) = \mathbf{V}^{(t)}(\omega) \mathbf{\Lambda}^{(t)}(\omega) \mathbf{U}^{(t)}(\omega)^*, \quad (4)$$

where $\mathbf{U}^{(t)}(\omega)$ and $\mathbf{V}^{(t)}(\omega)$ are frequency-dependent input and output unitary beam-forming matrices, respectively, and

$$\mathbf{\Lambda}^{(t)}(\omega) = \mathrm{diag}\left[ e^{\frac{1}{2}g_1^{(t)}(\omega)}, e^{\frac{1}{2}g_2^{(t)}(\omega)}, \ldots, e^{\frac{1}{2}g_D^{(t)}(\omega)} \right]. \quad (5)$$

Here, $\mathbf{g}^{(t)}(\omega) = \left(g_1^{(t)}(\omega), g_2^{(t)}(\omega), \ldots, g_D^{(t)}(\omega)\right)$ is a frequency-dependent vector of the logarithms of the eigenvalues of $\mathbf{M}^{(t)}(\omega)\mathbf{M}^{(t)}(\omega)^*$, which quantifies the overall MDL of a MIMO system.

In the MIMO system characterized by the random matrix $\mathbf{M}^{(t)}(\omega)$, the GDs of the modes are given by the eigenvalues of $j\mathbf{M}_\omega^{(t)}(\omega)\mathbf{M}^{(t)}(\omega)^*$, where $\mathbf{M}_\omega^{(t)}(\omega) = \mathrm{d}\mathbf{M}^{(t)}(\omega)/\mathrm{d}\omega$ [14]. In the absence of MDL, in an MMF with $K$ statistically identical sections, the GDs have a variance $\sigma_{\mathrm{gd}}^2 = K\sigma_\tau^2$, where $\sigma_\tau^2$ is the GD variance of an individual section [14][28]. Moreover, in the absence of MDL, the GDs are frequency-dependent, but all statistical properties of the GDs depend only on the number of modes and the overall GD STD $\sigma_{\mathrm{gd}} = \sqrt{K}\sigma_\tau$ (at least when chromatic dispersion is the same for all spatial modes). For an MMF with MDL, the statistical properties of the GDs are more complicated, and are outside the scope of this paper.

For the MDL at each single frequency, the MDL statistics depend only on the number of modes and on the square-root of the accumulated MDL variance $\xi$ via [15]:

$$\sigma_{\mathrm{mdl}} = \xi\sqrt{1 + \tfrac{1}{12}\xi^2} . \quad (6)$$

If an MMF comprises $K$ independent, statistically identical sections, each with MDL variance $\sigma_g^2$, we have $\xi = \sqrt{K}\sigma_g$. The MDL at each single frequency has these statistical properties, regardless of the GD STD $\sigma_{\mathrm{gd}}$.

### B. Frequency Dependence of MDL

The MDL given by the singular value decomposition (4) is frequency-dependent in general. In the special case that there is no modal dispersion, such that $\sigma_{\mathrm{gd}} = \sqrt{K}\sigma_\tau$ is equal to zero, the MDL is independent of frequency. Assuming nonzero $\sigma_{\mathrm{gd}}$, the correlation of the MDL at two frequencies depends on the frequency separation. If the frequency separation is small, the phase factors for the uncoupled modes appearing in (3) are similar, leading to similar MDL values at the two frequencies. If the frequency separation is large, the values of $\mathbf{M}^{(t)}(\omega)$ at the two frequencies are independent, leading to independent MDL at the two frequencies.

Considering the simplest case of two modes, Figure 1 illustrates the frequency dependence of MDL in the regimes of small and large GD spread, quantified by the GD STD $\sigma_{\mathrm{gd}}$. Over the frequency range shown, the gains $g_1^{(t)}(\omega)$ and $g_2^{(t)}(\omega)$ (and thus the MDL) vary slowly for small $\sigma_{\mathrm{gd}}$ and rapidly for large $\sigma_{\mathrm{gd}}$, as shown in Fig. 1(a) and (b), respectively. Assuming signals are launched into two orthogonal reference modes, the output powers (in logarithmic units) are

$$\log\left( \tfrac{1}{2}\left|M_{11}^{(t)}\right|^2 + \tfrac{1}{2}\left|M_{12}^{(t)}\right|^2 \right) \text{ and } \log\left( \tfrac{1}{2}\left|M_{21}^{(t)}\right|^2 + \tfrac{1}{2}\left|M_{22}^{(t)}\right|^2 \right),$$

---

[1]Throughout this paper, "modes" include both polarization and spatial degrees of freedom. For example, For example, the two-mode case can describe the two polarization modes in single-mode fiber.

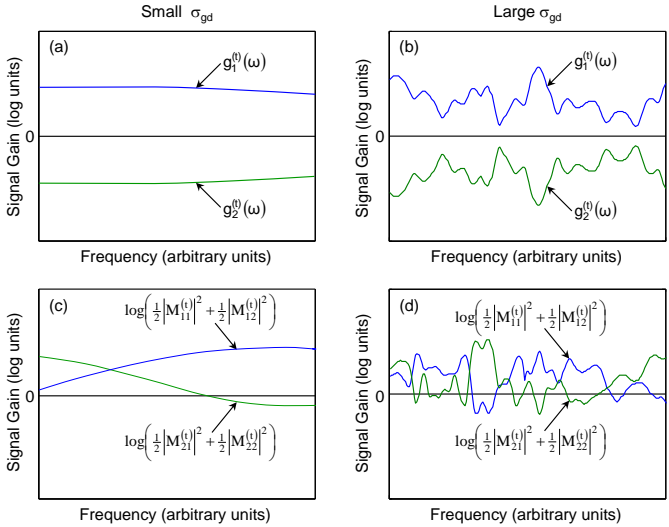

Fig. 1 Frequency dependence of the MDL in a two-mode fiber for (a) small $\sigma_{gd}$ and (b) large $\sigma_{gd}$, where $\sigma_{gd}$ is the STD of GD. Output powers of signals launched into two orthogonal reference modes for (c) small $\sigma_{gd}$ and (d) large $\sigma_{gd}$.

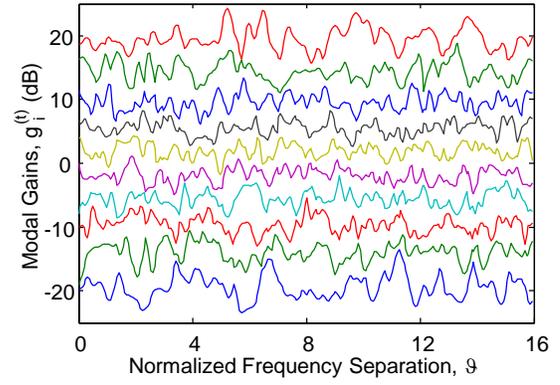

Fig. 2 Modal gains $g_i^{(t)}$, $i = 1, \ldots, D$, as a function of normalized frequency separation for an MMF with $D = 10$ modes.

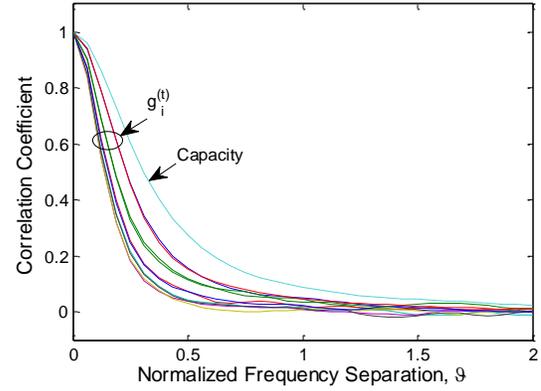

Fig. 3 Correlation coefficients of modal gains $g_i^{(t)}$, $i = 1, \ldots, D$, as a function of normalized frequency separation for an MMF with $D = 10$ modes. The correlation coefficient of the average channel capacity, assuming an SNR of 20 dB and assuming CSI is not available at the transmitter, is also shown.

respectively. Over frequency, these output powers vary slowly for small $\sigma_{gd}$ and rapidly for large $\sigma_{gd}$, as shown in Fig. 1(c) and (d), respectively. For MDM signals spanning the frequency range shown, Figs. 1(c) and (d) would correspond to regimes of low diversity order and moderate-to-high diversity order, respectively.

The correlation properties of MDL should depend on the normalized frequency separation $\vartheta = \Delta\omega\,\sigma_{gd}/2\pi$, where $\Delta\omega$ is the angular frequency separation. For small normalized frequency separation, $\vartheta \ll 1$, the MDLs at the two frequencies are identical, while for large normalized frequency separation, $\vartheta \gg 1$, the MDLs at the two frequencies are independent. The coherence bandwidth of MDL should be of the same order as the reciprocal of the overall STD of GD, $1/\sigma_{gd}$; hence, the normalized coherence bandwidth should be of order unity.

Figure 2 shows simulations of the gain vector $\mathbf{g}^{(t)}(\omega)$ defined in (5) as a function of normalized frequency separation $\vartheta$. The MMF has $D = 10$ modes and an accumulated MDL of $\xi = 10$ dB. The MMF comprises $K = 256$ statistically identical sections, as in [15]. The gain vector in each section $\mathbf{g}^{(k)}$ is the same as in [15]. The GD vector $\mathbf{\tau}^{(k)}$ in each section is generated as a Gaussian random vector whose entries sum to zero, using the method described in the Appendix of [15]. Each curve in Fig. 2 corresponds to one of the elements of the vector $\mathbf{g}^{(t)}(\omega)$ as a function of normalized frequency separation $\vartheta$. The $x$-axis of Fig. 2 is the normalized frequency separation with respect to the first frequency.

Figure 2 illustrates how the correlation of the MDL depends on frequency separation, similar to Figs. 1(a) and (b). The gain of each mode is a smooth, continuous curve, so each modal gain is highly correlated for small frequency separations. Conversely, each modal gain is uncorrelated for large frequency separations. Figure 2 also shows that the highest and lowest modal gains are subject to larger variations than the intermediate modal gains, consistent to the theory of [14][15], in which the outer peaks of the probability density function exhibit a larger spread then the inner peaks.

Figure 3 shows the correlation coefficients of the elements of the modal gain vector $\mathbf{g}^{(t)}(\omega)$ as a function of normalized frequency separation. The simulation parameters are the same as in Fig. 2, but the correlation coefficients are calculated with 23,000 realizations of modal gain curves, each similar to Fig. 2. In Fig. 3, the correlation coefficient is calculated for each gain coefficient after conversion to a decibel scale. In Fig. 3, the ten curves are observed to cluster into five pairs, which are for the gain coefficients $g_i^{(t)}$ and $g_{D-i+1}^{(t)}$, $i = 1, \ldots, 5$. The correlation coefficients are observed to decrease with an increase of $i$. Referring to Fig. 2, the highest and lowest two curves ($i = 1$) exhibit the largest (and similar) correlation over frequency, while the middle two curves ($i = 5$) exhibit the smallest (and similar) correlation over frequency.

In Fig. 3, it is difficult to uniquely define a single coherence bandwidth for all the modal gains, because they decay at different rates, and do not decay fully to zero at large normalized frequency separation (this may arise, at least in part, from numerical errors). Considering the highest and lowest gains with the largest correlations, the normalized one-sided coherence bandwidths are 0.25 or 0.67 for correlation coefficients of 50% or 10%, respectively. At 10% correlation coefficient, the normalized one-sided coherence bandwidth ranges from 0.32 to 0.67 for the different gain coefficients. At a normalized frequency separation of unity, the correlation coefficients range from 0 to 4.7% for the different gain coefficients.

### III. CHANNEL CAPACITY AND FREQUENCY DIVERSITY

Figures 1-3 demonstrate that the modal gains are frequency-dependent, and are strongly correlated only over a finite coherence bandwidth. If MDM signals occupy a bandwidth far larger than the coherence bandwidth, the outage channel capacity should approach the ensemble average channel capacity of the channel due to statistical averaging.

#### A. Outage and Average Channel Capacities

At any single frequency, the gain vector $\mathbf{g}^{(t)}(\omega)$ given by (5), obtained by the singular value decomposition (4), is a random vector having the same statistical properties as a zero-trace Gaussian unitary ensemble, assuming a system with practical MDL values [15]. The channel capacity is also a random variable, and it depends on whether or not channel state information (CSI) is available at the transmitter [15] [16]. Assuming CSI is not available, given a realization of the gain vector $\mathbf{g}^{(t)}(\omega)$, at a single frequency, the channel capacity[2] is:

$$C = \sum_{i=1}^{D} \log_2\left[1 + \frac{\chi}{D}\exp\left(g_i^{(t)}\right)\right], \quad (7)$$

where $\chi$ is the ratio of the transmitted power (total over all $D$ modes) to the received noise power (per mode) [15]. The SNR[3] is the product of $\chi$ and the average gain $E\left\{\exp\left(g_i^{(t)}\right)\right\}$, where $E\{\ \}$ denotes expectation. The channel capacity over the signal bandwidth is just the average of (7) over the signal bandwidth. The capacity (7) assumes that the output noises in the principal modes are independent and identically distributed (i.i.d). This assumption is justified theoretically and verified numerically in [15].

Figure 4 shows the simulated distribution of the channel capacity of a MDM system with $D = 10$ modes at an SNR of 20 dB, corresponding to an SNR per mode of 10 dB, assuming CSI is not available at the transmitter. All parameters are the same as in Fig. 3. The distribution in Fig. 4 is constructed using about 5,900,000 channel capacity values.

In Fig. 4, we observe that the average channel capacity, near the peak of the distribution, is about 17.2 b/s/Hz, while the capacity for $10^{-3}$ outage probability is about 14.3 b/s/Hz. At any single frequency, the channel capacity has the same distribution as that in Fig. 4.

#### B. Correlation Coefficient of Channel Capacity

We recall that Fig. 3 shows the correlation coefficients of the modal gains vs. normalized frequency separation, for an MMF with $D = 10$ modes, illustrating how the gains at nearby frequencies are highly correlated. Figure 3 also shows the correlation coefficient of the average channel capacity. The capacity is computed as in Fig. 4, assuming an SNR of 20 dB, and assuming no CSI is available at the transmitter, so equal power is allocated to all modes. The normalized one-sided coherence bandwidths of the channel capacity are 0.31 and 0.92 for correlation coefficients of 50% and 10%, respectively. A normalized frequency separation of unity gives a correlation coefficient of 8.8%.

---

[2] Throughout this paper, channel capacity is computed per unit bandwidth, and thus has units of b/s/Hz.
[3] As in [15], following the literature on MIMO wireless systems, the SNR is defined as the received signal power (total over all $D$ modes) divided by the received noise power (per mode).

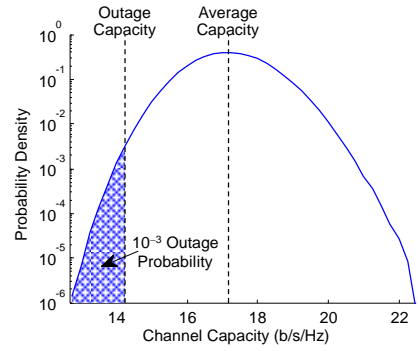

Fig. 4 Distribution of channel capacity at a single frequency for a MDM system with $D = 10$ modes at an SNR of 20 dB, assuming CSI is not available at the transmitter. The capacity at $10^{-3}$ outage probability is indicated.

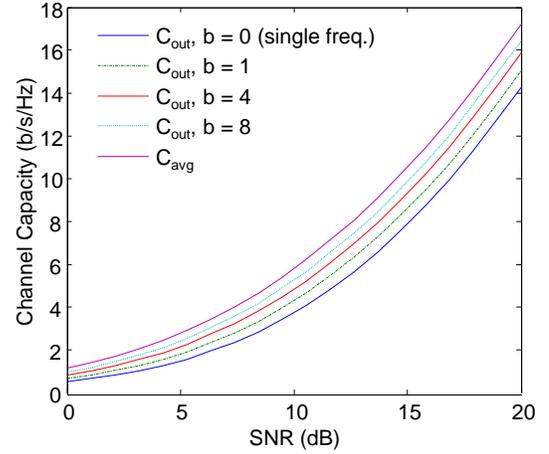

Fig. 5 Outage channel capacity at $10^{-3}$ outage probability vs. SNR for various normalized bandwidths, for an MDM system with $D = 10$ modes, assuming CSI is not available at the transmitter. The average channel capacity is shown for comparison.

#### C. Frequency Diversity and Diversity Order

If MDM signals occupy a bandwidth much greater than the coherence bandwidth of the capacity, statistical averaging over frequency should cause the outage channel capacity to approach the average channel capacity. Figure 5 shows the outage capacity at as a function of SNR, for signals occupying different bandwidths. All parameters are as in Figs. 3 and 4, i.e., $D = 10$ modes are used, CSI is not available, and the outage probability is $10^{-3}$. A normalized signal bandwidth is defined as $b = B_{\text{sig}}\sigma_{\text{gd}}$, where $B_{\text{sig}}$ is the signal bandwidth (measured in Hz). In Fig. 5, signals occupy normalized bandwidths from $b = 0$ (a single frequency) to $b = 8$. The average capacity is shown for comparison. Figure 5 shows that as the normalized bandwidth increases, the outage capacity does approach the average capacity.

Statistical averaging over frequency is a consequence of the law of large numbers [29]. Consider two OFDM subchannels at frequencies whose separation far exceeds the coherence bandwidth of the capacity, e.g., at two frequencies well-separated in Fig. 1(b). Suppose the subchannels have capacities $C_1$ and $C_2$; these are independent random variables following a common distribution (e.g., that in Fig. 4). A channel comprising the two subchannels has an overall capacity $\frac{1}{2}(C_1 + C_2)$, as the capacity is computed on a per-unit-frequency basis. If each subchannel capacity has variance $\sigma_C^2$, the overall channel capacity has variance $\frac{1}{2}\sigma_C^2$, i.e. it is reduced by a factor of two. The channel comprising two i.i.d. subchannels has a diversity order of two. More generally, given a channel spanning a finite bandwidth $B_{\text{sig}}$, we define the *diversity order* in terms of a reduction of the variance of capacity: a diversity order equal to $F_D$ corresponds to a reduction of the variance of capacity from $\sigma_C^2$ to $\sigma_C^2/F_D$. With this definition, the diversity order may be any real number not smaller than unity.

It would be useful to be able to estimate the diversity order $F_D$ directly from the statistics of the frequency-dependent gain vector $\mathbf{g}^{(t)}(\omega)$, rather than having to compute the frequency-dependent capacity and characterize its statistics. In the Appendix, a procedure is described for computing the diversity order directly from the frequency correlation coefficients of the channel capacity using principal component analysis.

Numerical simulations of MDM systems similar to those in Fig. 5 have been performed, with number of modes $D = 10$ and normalized bandwidth $b$ ranging from 0 to 16. As the diversity order $F_D$ computed using (10) varies, the distribution of the channel capacity is found to retain approximately the same shape as that of Fig. 4, but the variance is reduced from $\sigma_C^2$ to approximately $\sigma_C^2/F_D$. The mean channel capacity is found not to change with diversity order. The outage capacity as a function diversity order is found to follow

$$C_{\text{out},F_D} \approx C_{\text{avg}} - \frac{1}{\sqrt{F_D}}\left(C_{\text{avg}} - C_{\text{out},1}\right), \quad (8)$$

where $C_{\text{out},1}$ is the single-frequency outage capacity following the distribution in Fig. 4. The relationship (8) is found to be independent of the outage probability, provided the outage capacities $C_{\text{out},1}$ and $C_{\text{out},F_D}$ refer to the same outage probability. The relationship (8) is found to be valid as the shape of the distribution of capacity deviates from that shown in Fig. 4. If the capacity distribution is assumed to be Gaussian, as in [16], the relationship between outage capacity and diversity order $F_D$ at any particular outage probability can be computed analytically. However, the distribution of capacity in Fig. 4 is observed to deviate noticeably from a Gaussian distribution, e.g., it is slightly asymmetric. This non-

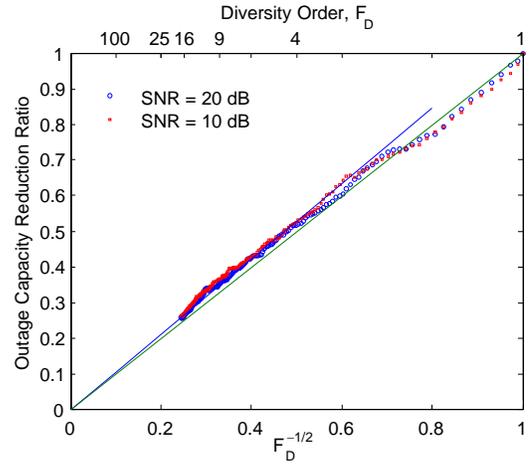

Fig. 6 The outage capacity reduction ratio, given by (9), as a function of $1/\sqrt{F_D}$, where $F_D$ is the diversity order, given by (10). The green and blue lines have slopes of 1 (theoretical slope) and 1.06 (best-fit slope), respectively. The MDM system uses $D = 10$ modes, and CSI is not available at the transmitter.

Gaussianity is consistently observed at all SNRs and all diversity orders.

Figure 6 shows the outage capacity reduction ratio, defined as

$$\frac{C_{\text{avg}} - C_{\text{out},F_D}}{C_{\text{avg}} - C_{\text{out},1}} \quad (9)$$

as a function of $1/\sqrt{F_D}$, where $F_D$ is the diversity order computed using (10) from the correlation coefficients of the channel capacity $C$ shown in Fig. 3. The simulation parameters used for Fig. 6 are the same as those of Figs. 3 and 5, i.e., the MDM system uses $D = 10$ modes, and CSI is not available at the transmitter. Values of the diversity order $F_D$ are only computed for SNR = 20 dB, as in Fig. 3, but Fig. 6 shows values of (9) computed at SNR = 10 and 20 dB, illustrating that the diversity order $F_D$ is valid over a range of SNR values. The correlation coefficients in Fig. 3 are subject to numerical error, as they never go to zero even for large frequency separations. To limit numerical error, diversity orders in Fig. 6 are computed only using values of the correlation coefficients from Fig. 3 that are larger than 1%.

Based on (8), the outage capacity reduction ratio (9) should approximately equal to $1/\sqrt{F_D}$, and the plots in Fig. 6 should be straight lines with unit slope. In Fig. 6, the best-fit slope is found to be 1.06. Figure 6 clearly shows that $C_{\text{avg}} - C_{\text{out},F_D}$ approaches zero as the diversity order $F_D$ increases. The observed dependence of the difference between average and outage capacities on $1/\sqrt{F_D}$ is a direct consequence of the law of large numbers [29]. The outage and average capacities converge slowly with an increase in diversity order $F_D$. The diversity order $F_D$ must be four to decrease the capacity difference to half that without diversity, and must be 100 to decrease the difference to 10% of that without diversity.

## IV. DISCUSSION

In MDM systems using coherent detection, modal dispersion does not fundamentally degrade performance, but does affect the complexity of signal processing required for equalization and spatial demultiplexing [14]. By contrast, MDL can fundamentally degrade MDM system performance [15], particularly when CSI is not available at the transmitter. Hence, it can be advantageous to design a transmission system with sufficient modal dispersion to provide the frequency diversity needed to mitigate MDL. This work, together with [14][15], provides a basis for transmission system design to counter MDL.

The ratio of outage to average capacities, $C_{\text{out},1}/C_{\text{avg}}$, decreases with an increase of MDL or a reduction of SNR. In the limit of a very high SNR and an MDL smaller than the SNR, the channel capacity without CSI (7) is approximately equal to $D\log_2 \chi/D + \log_2 e \sum_{i=1}^{D} g_i^{(t)} = D\log_2 \chi/D$, which is independent of the frequency-dependent gain vector $\mathbf{g}^{(t)}(\omega)$. The average SNR needs to be large enough that even the weakest mode has sufficiently high SNR. At high SNR, the channel capacity is independent of $\mathbf{g}^{(t)}(\omega)$ even for a system with CSI.

In Fig. 3, the modal gains are observed to have smaller coherence bandwidths than the channel capacity at typical SNR values. At low SNR, as shown in [15], the channel capacity is proportional to the overall received power. With large MDL, the coherence bandwidth of the capacity is determined by the modes having the largest gain. At low SNR, the channel capacity has smaller coherence bandwidth than the modal gains.

While not yet established, in the case of spatial-mode-dependent chromatic dispersion, the higher-order frequency dependence of the GD statistics may be modified slightly [14]. The general approach presented here should remain valid. The STD of GD becomes a frequency-dependent $\sigma_{gd}(\omega)$ to include the effect of spatial-mode-dependent chromatic dispersion. The normalized frequency separation may be modified to $\vartheta = (2\pi)^{-1} \int_0^{\Delta\omega} \sigma_{gd}(\omega_0 + s) ds$, where $\omega_0$ is the reference frequency. The normalized bandwidth $b$ is always the normalized frequency separation between the lowest and highest frequencies.

In mobile wireless systems, the frequency diversity order [30] has been estimated as $\lceil B_{\text{sig}}/B_{\text{coh}} \rceil$, where $B_{\text{coh}}$ is a coherence bandwidth of the channel frequency response, and where $\lceil x \rceil$ denotes the smallest integer greater than or equal to $x$. Here, for large $B_{\text{sig}}$, $\lceil B_{\text{sig}}/B_{\text{coh}} \rceil$ is found to be close to the correct diversity order $F_D$, i.e., close to the observed reduction of the variance of capacity, but for small $B_{\text{sig}}$, $\lceil B_{\text{sig}}/B_{\text{coh}} \rceil$ tends to underestimate the diversity order. This is not surprising; for example, when $B_{\text{sig}} = B_{\text{coh}}$, $\lceil B_{\text{sig}}/B_{\text{coh}} \rceil$ yields a diversity order of one, whereas the correct diversity order should lie between one and two (for example, $F_D = 1.89$ in Fig. 6).

## V. CONCLUSION

In the strong-coupling regime, the frequency dependence of MDL in an MMF has a coherence bandwidth inversely proportional to the STD of GD. If an MDM signal occupies a bandwidth far larger than the coherence bandwidth, because of statistical averaging, the outage channel capacity approaches the average channel capacity. The difference between the average and outage channel capacities decreases with the square-root of a diversity order. The diversity order can be computed using the frequency correlation coefficients of the modal gains based on principal component analysis. If the signal has a bandwidth far larger than the coherence bandwidth, the diversity order is the ratio of the signal bandwidth to the coherence bandwidth.

## APPENDIX

This Appendix describes procedures for computing the diversity order $F_D$ directly from the frequency correlation coefficients of the modal gains using principal component analysis [31]. Principal component analysis is similar to either the continuous or discrete Karhunen–Loève transform [32][33], which yield similar results.

Given $\mathbf{R}$, the covariance or correlation matrix of a vector of random variables, the number of independent components can be found by the eigenvalue decomposition of $\mathbf{R}$ [31]. In the present context, the number of independent components corresponds to the diversity order $F_D$. While conceptually similar, the number of independent components $F_D$ may be defined in terms of:

1. The number of non-zero eigenvalues.
2. The number of eigenvalues up to certain fraction (e.g., 1%) of the largest eigenvalue.
3. The sum of all eigenvalues, scaled by the largest eigenvalue. If the eigenvalues are $\lambda_1, \lambda_2, \ldots$ with $\lambda_1$ largest, the number of independent components is

$$F_D = \sum_k \frac{\lambda_k}{\lambda_1}. \quad (10)$$

The first definition is the same as the rank of the matrix [34][35]. In both the first and second definitions, the number of independent components is an integer [31], similar to the definition of diversity order in [30][34][35]. In the third definition, the number of independent components may be a real number greater than or equal to unity. As suggested by (9), the diversity order $F_D$ need not be integer-valued, so we employ the third definition here.

The diversity order can be equated to the number of independent components of a correlation matrix

$$\mathbf{R} = \begin{bmatrix} c_0 & c_1 & c_2 & \cdots \\ c_1 & c_0 & c_1 & \cdots \\ c_2 & c_1 & c_0 & \cdots \\ \vdots & \vdots & \vdots & \ddots \end{bmatrix},$$

where each row of the matrix corresponds to a correlation coefficient, sampled uniformly at different values of the frequency difference. Such a correlation matrix $\mathbf{R}$ is a symmetric Toeplitz matrix given by a sequence $\{c_0, c_1, c_2, \ldots\}$ [36], which is fully specified by the frequency correlation coefficients (e.g., those in Fig. 3).

In (10), the summation $\sum_k \lambda_k$ is the trace of $\mathbf{R}$, and is equal to $Nc_0$ where $N$ is the dimension of $\mathbf{R}$. If OFDM signals are used, $N$ may be interpreted as the number of subchannels. If the dimension of a Toepltiz matrix is very large, its eigenvalues are given by the Fourier transform of the sequence $\{\ldots, c_2, c_1, c_0, c_1, c_2, \ldots\}$ [37]. For example, if $N$ is an odd number, the largest eigenvalue is $\lambda_1 = c_0 + 2\sum_{k=1}^{(N-1)/2} c_k$. In Fig. 3, the correlation coefficients are observed to be small for large $k$. From the theory of large Toeplitz matrices, the diversity order (10) is always proportional to $N$ and always directly proportional to the signal bandwidth if the number of OFDM subchannels $N$ is very large and $c_{N/2}$ approaches zero. The coherence bandwidth may be defined as

$$B_{\text{coh}} = \lim_{B_{\text{sig}} \to \infty} \frac{B_{\text{sig}}}{F_D},$$

which yields

$$B_{\text{coh}} = \lim_{B_{\text{sig}} \to \infty} \frac{B_{\text{sig}}}{N} \left(1 + \frac{2}{c_0} \sum_{k=1}^{N/2} c_k \right), \quad (11)$$

where $B_{sig}/N$ is equal to the subchannel spacing in the case of OFDM signals. If the correlation coefficient versus frequency is $c(f)$, the coherence bandwidth (11) is just $\int_{-\infty}^{+\infty} c(f) df$ with $\lim_{f \to \pm\infty} c(f) = 0$ and $c(0) = 1$, showing obviously that (11) yields a two-sided coherence bandwidth.

In this paper, the diversity order $F_D$ is always computed using (10) by finding the eigenvalues of the Toeplitz matrix numerically. For zero-mean stationary random process, the covariance and correlation matrices are equivalent and yield identical results.

## REFERENCES

[1] P. K. Pepeljugoski and D. M. Kuchta, "Design of optical communications data links," *IBM J. Res. Dev.*, vol. 47, no. 2-3, pp. 223–237, 2003.
[2] R. E. Freund, C.-A. Bunge, N. N. Ledentsov, D. Molin, and C. Caspar, "High-speed transmission in multimode fibers," *J. Lightw. Technol.*, vol. 28, no. 4, pp. 569-586, 2010.
[3] Y. Koike and S. Takahashi, "Plastic optical fibers: technologies and communication links," in *Optical Fiber Telecommunications VB: Systems and Networks*, I. P. Kaminow, T. Li and A. E. Willner eds. San Diego: Elsevier Academic, 2008.
[4] H. R. Stuart, "Dispersive multiplexing in multimode optical fiber," *Science*, vol. 289, no. 5477, pp. 281-283, 2000.
[5] A. Tarighat, R. C. J. Hsu, A. Shah, A. H. Sayed, and B. Jalali, "Fundamentals and challenges of optical multiple-input multiple-output multimode fiber links," *IEEE Commun. Mag.*, pp. 57-63, May 2007.
[6] A. R. Shah, R. C. J. Hsu, A. Tarighat, A. H. Sayed, and B. Jalali, "Coherent optical MIMO (COMIMO)," *J. Lightw. Technol.*, vol. 23, no. 8, pp. 2410-2419, 2005.
[7] M. Nazarathy and A. Agmon, "Coherent transmission direct detection MIMO over short-range optical interconnects and passive optical networks," *J. Lightw. Technol.*, vol. 26, no. 14, pp. 2037–2045, 2008.
[8] A. Al Amni, A. Li, S. Chen, X. Chen, G. Gao, and W. Shieh, "Dual-LP$_{11}$ mode $4 \times 4$ MIMO-OFDM transmission over a two-mode fiber," *Opt. Express*, vol. 19, no. 17, pp. 16672–16678, 2011.
[9] S. Randel, R. Ryf, A. Sierra, P. Winzer, A. H. Gnuack, C. Bolle, R.-J. Essiambre, D. W. Peckham, A. McCurdy, and R. Lingle, "$6 \times 56$-Gb/s mode-division multiplexed transmission over 33-km few-mode fiber enabled by $6 \times 6$ MIMO equalization," *Opt. Express*, vol. 19, no. 17, pp. 16697–16707, 2011.
[10] C. Koebele, M. Salsi, D. Sperti, P. Tran, P. Brindel, H. Margoyan, S. Bigo, A. Boutin, F. Verluise, P. Sillard, M. Bigot-Astruc, L. Provost, F. Cerou, and G. Charlet, "Two mode transmission at $2 \times 100$ Gb/s, over 40 km-long prototype few-mode fiber, using LCOS based mode multiplexer and demultiplexer," *Opt. Express*, vol. 19, no. 17, pp. 16593–16600, 2011.
[11] D. Gloge, "Weakly guiding fibers," *Appl. Opt.,* vol. 10, no. 10, pp. 2252-2258, 1970.
[12] D. Gloge, "Optical power flow in multimode fibers," *Bell System Tech. J.,* vol. 51, pp. 1767-1780, 1972.
[13] R. Olshansky, "Mode-coupling effects in graded-index optical fibers," *App. Opt.*, vol. 14, no. 4, pp. 935-945, 1975.
[14] K.-P. Ho and J. M. Kahn, "Statistics of group delays in multimode fiber with strong mode coupling," to be published in *J. Lightw. Technol.*, http://arxiv.org/abs/1104.4527.
[15] K.-P. Ho and J. M. Kahn, "Mode-dependent loss and gain: Statistics and effect on mode-division multiplexing," *Opt. Express*, vol. 19, no. 17, pp. 16612-16635, 2011.
[16] P. J. Winzer and G. J. Foschini, "MIMO capacities and outage probabilities in spatially multiplexed optical transport systems," *Opt. Express*, vol. 19, no. 17, pp. 16680–16696, 2011.
[17] T. S. Rappaport, *Wireless Communications: Principles & Practice,* Upper Saddle River, NJ: Prentice Hall, 2002, Sec. 5.5.1.2.
[18] H. Sari, G. Karam, and I. Jeanclaude, "Transmission techniques for digital terrestrial TV broadcasting," *IEEE Commun. Mag.,* pp. 100-109, Feb. 1995.
[19] W. Y. Zou and Y. Wu, "COFDM: An overview," *IEEE Trans. Broadcast.,* vol. 41, no. 1, pp. 1-8, 1995.
[20] E. Lindskog and A. Paulraj, "A transmit diversity scheme for delay spreads channel," in *IEEE ICC 2000*, pp. 207-311.
[21] K. F. Lee and D. B. Williams, "A space-frequency transmitter diversity techniques for OFDM systems," in *IEEE GlobeCom 2000*, vol. 3, pp. 1473-1477.
[22] N. Al-Dhahir, "Single-carrier frequency-domain equalization for space-time block-coded transmissions over frequency-selective fading channels," *IEEE Commun. Lett.,* vol. 5, no. 7, pp. 304-306, 2001.
[23] X. Zhu and R. D. Murch, "Layered space-frequency equalization in a single-carrier MIMO system for frequency-selective channels," *IEEE Trans. Wireless Commun.,* vol. 3, pp. 701-708, 2004.
[24] M. Karlsson, "Probability density functions of the differential group delay in optical fiber communication systems," *J. Lightw. Technol.*, vol. 19, no. 3, pp. 324-331, 2001.
[25] J. P. Gordon and H. Kogelnik, "PMD fundamentals: Polarization mode dispersion in optical fibers," *Proc. Natl. Acad. Sci.*, vo. 97, no. 9, pp. 4541-4550, 2000.
[26] A. M. Tulino and S. Verdú, "Random Matrix Theory and Wireless Communications," *Foundation and Trends in Communications and Information Theory,* vol. 1, no. 1, pp. 1-182, 2004.
[27] D. Tse and P. Viswanath, *Fundamentals of Wireless Communication*, Cambridge UK: Cambridge Univ. Press, 2005.
[28] M. B. Shemirani, W. Mao, R. A Panicker, and J. M. Kahn, "Principal modes in graded-index multimode fiber in presence of spatial- and polarization-mode coupling," *J. Lightw. Technol.*, vol. 27, no. 10, pp. 1248-1261, 2009.
[29] G. R. Grimmett and D. R. Stirzaker, *Probability and Random Processes*, 2nd ed. Oxford, UK: Oxford Univ. Press, 1992, Sec. 5.10.3.

[30] K. Liu, T. Kadous, and A. K. Sayeed, "Orthogonal time-frequency signaling over doubly dispersive channels," *IEEE Trans. Info. Theory,* vol. 50, no. 11, pp. 2583-2602, 2004.
[31] I. T. Jolliffe, *Principal Component Analysis*, 2nd ed., New York: Springer, 2002.
[32] W. B. Davenport and W. L. Root, *An Introduction to the Theory of Random Signals and Noise,* New York: Wiley-IEEE Press, 1987, Sec. 6-4.
[33] M. Vetterli and J. Kovačević, *Wavelets and Subband Coding,* Englewood Cliffs, NJ: Prentice Hall, 1995, Sec. 7.1.1.
[34] V. Tarokh, H. Jafarkhani, and A. R. Calderbank, "Space-time block codes from orthogonal designs," *IEEE Trans. Info. Theory,* vol. 45, no. 5, pp. 1456-1467, 1999.
[35] X. Ma and G. B. Giannakis, "Maximum-diversity transmissions over doubly selective wireless channels," *IEEE Trans. Info. Theory,* vol. 49, no. 7, pp. 1832-1840, 2003.
[36] R. M. Gray, "Toeplitz and Circulant Matrices: A Review," *Foundation and Trends in Communications and Information Theory,* vol. 2, no. 3, pp. 155-239, 2006.
[37] R. M. Gray, "On the asymptotic eigenvalue distribution of Toeplitz matrices," *IEEE Trans. Info. Theory,* vol. IT-18, no. 6, pp. 725-730, 1972.